# A Scintillating Fiber Dosimeter for Radiology and Brachytherapy with photodiode readout

Florbela Rêgo[a)], and Maria da Conceição Abreu

*Laboratório de Instrumentação e Física Experimental de Partículas, Lisboa, Portugal*

Luis Peralta

*Universidade de Lisboa, Faculdade de Ciências, Departamento de Física, Portugal and*

*Laboratório de Instrumentação e Física Experimental de Partículas, Lisboa, Portugal*

**Abstract**

**Purpose:** For more than a decade that plastic optical fiber based dosimeters have been developed for medical applications. The feasibility of dosimeters using optical fibers that are almost Cherenkov light free has been demonstrated in some prototypes, particularly suitable for photon high-energy beams. In the energy range up to a few hundred keV, where the production of Cherenkov light by secondary electrons is negligible or small, the largest source of background are the fluorescence mechanisms.

**Methods:** In recent years our group has developed an optical fiber dosimeter with photodiode readout named DosFib, which has small energy dependence in the range below 100 keV relevant for radiology. Photodiodes are robust photodetectors, presenting good stability over time and enough sensitivity to allow the use of an electrometer as a measuring device without extra electronics.

**Results:** In-vitro tests using a High Dose Rate $^{192}$Ir source have demonstrated its suitability for brachytherapy applications using this important radioactive source. Attenuation curves have been obtained in water with the DosFib dosimeter and an ionization chamber and the depth dose profiles shown to be identical.

**Conclusions:** This new dosimeter can thus be calibrated and used in clinical measurements with good competitive advantages over other dosimeters.

## I. INTRODUCTION

Plastic scintillators possess favorable characteristics to be used as dosimeters in radiology or other applications using low energy photons. Plastics have good radiological equivalence with water and can be made with dimensions of the order of a millimeter. Plastic scintillating optical fibers are an example of this type of material. Scintillation light is emitted by the dopant of the plastic bulk material when irradiated by ionizing radiation. The amount of light emitted is proportional to the energy deposited, at least for low ionizing density radiation like X or gamma photons.[1] Unfortunately, fluorescent light is also produced by the fiber bulk material and, as shown in earlier works[2] as well as in this one, the fluorescent light emitted from the bulk material introduces a dependence of the output signal on the incident radiation energy. Other well-known source of background is the Cherenkov light emitted by electrons.[3] Cherenkov light is generated in a medium of refraction index $n$ when electrons travel faster than $c/n$, where $c$ is the speed of light in vacuum. For optical fibers, the signal from Cherenkov light is highly dependent on the fiber-beam relative configuration and in particular the angle between the electron direction and the fiber axis.[4] Moreover, the amount of deposited energy needed to produce Cherenkov photons is much less than the amount needed to produce scintillation photons of the same wavelength,[5] introducing an energy dependence on the output signal. This subject has been extensively studied in the literature[3,4,6-12] and several solutions have been proposed to overcome the Cherenkov light influence in the final signal.[3,7,13] These solutions have particular relevance for the high energy beams used in external radiotherapy. In radiology the energy beam is below the Cherenkov threshold and the effect is not present. For brachytherapy applications that may use sources with intermediate energies, in the hundred keV range, some Cherenkov light will be produced.

Most of the scintillator based dosimeters developed over the years[3,7,14-16] use complex readout systems to avoid the Cherenkov light influence in the output signal. For medium/low energy the contribution due to Cherenkov light will be small or inexistent and the readout system can be greatly simplified. If brachytherapy High Dose Rate sources or high current radiology sources are used, then, the signal produced in small scintillator is sufficient to be detected by a Si-Pin photodiode or other modern solid-state photodetector avoiding the use of a PMT. This last aspect is important for future developments towards a portable device that can be used by the staff executing clinical procedures because, as pointed out by the International Commission on Radiological Protection, "*...many accidents could have been prevented if staff had had functional monitoring equipment and paid attention to the results* ".[17]

Bearing these aspects in mind, a dosimeter based on a scintillating plastic optical fiber with photodiode readout, named DosFib, was developed for diagnostic radiology and brachytherapy using high dose rate $^{192}$Ir sources. The dosimeter has good tissue equivalence, provides real time dose control, doesn't need high voltage and has good linearity with dose. The dosimeter has good radiation hardness, giving a large lifespan to the device. The optical fiber dosimeter with photodiode is almost temperature independent in a wide range, which is a competitive advantage over other readout devices.

## II. MATERIALS AND METHODS

75  **II.A. Dosimeter component selection**

*II.A.1 Photodetector selection*

Traditionally, scintillator dosimetry used PMT's as readout photodetector.[2,11,18,19] They have the advantage of signal high gain, but also present several disadvantages. They are noisy devices, with typical dark current in the range 1 to 10 nA for typical High Voltage (HV) bias values[*]. Since PMT needs HV to work they're not practical, for safety reasons, to make a portable system. The gain is highly dependent on the applied HV and, as a consequence, calibration constants are also dependent on the HV. Finally, the PMT is an expensive and mechanically fragile piece of equipment. As an alternative to PMT, several solid-state photodectors with high gain are nowadays available. On what concerns dosimetry, in many applications the light signal produced in the scintillator is high enough to be detected with a Si-PIN photodiode. These devices are robust, cheap and can be operated without any bias voltage. In this case the dark current can be as low as 5 pA[†], therefore, a thousand fold smaller than the typical PMT dark current. The photodiode can thus be DC coupled to the electrometer, greatly simplifying the output circuit. Furthermore, typical photo-sensitivities curves are in the range 300-1000 nm with maximum values in the near infrared[‡], making them suitable for the detection of a wide selection of scintillators.

For this study four different Si-PIN photodiodes from Hamamatsu, S9195, S1722-02 and S9075, S5344 (www.hamamatsu.com) have been tested. For the photodiodes tests a special setup was devised (figure

---

[*] Photomultiplier tubes, Basics and Applications, Hamamatsu Photonics K.K.
http://sales.hamamatsu.com/assets/applications/ETD/pmt_handbook_complete.pdf

[†] http://sales.hamamatsu.com/assets/pdf/parts_S/s1722-02_kpin1045e05.pdf
[‡] http://sales.hamamatsu.com/assets/pdf/parts_S/S9195.pdf

95    1). A 30 cm long, 2 mm in diameter scintillating fiber (always the same one) was placed inside a black light-tight perspex box. The box was shielded on the side by a Pb-Fe sandwich 2+2 mm in thickness, which prevents the fiber irradiation except for 15 mm due to a hole drilled in the box and shielding. Inside the test-box the fiber was placed on a groove opened on a black-perspex bed, ensuring complete placement reproducibility. At the end of the fiber a PVC holder ensured the coupling to the photodiode.
100   Good optical coupling between the fiber and the photodiode window was made using BC-630 optical grease from Saint-Gobain. The photodiode readout was made by a Standard Imaging MAX4000 electrometer. The optical fiber was irradiated with an X-ray beam produced by a Philips PW2184/00 tube with tungsten anode. The bench-table was leveled so the fiber stood at the center of the X-ray beam.

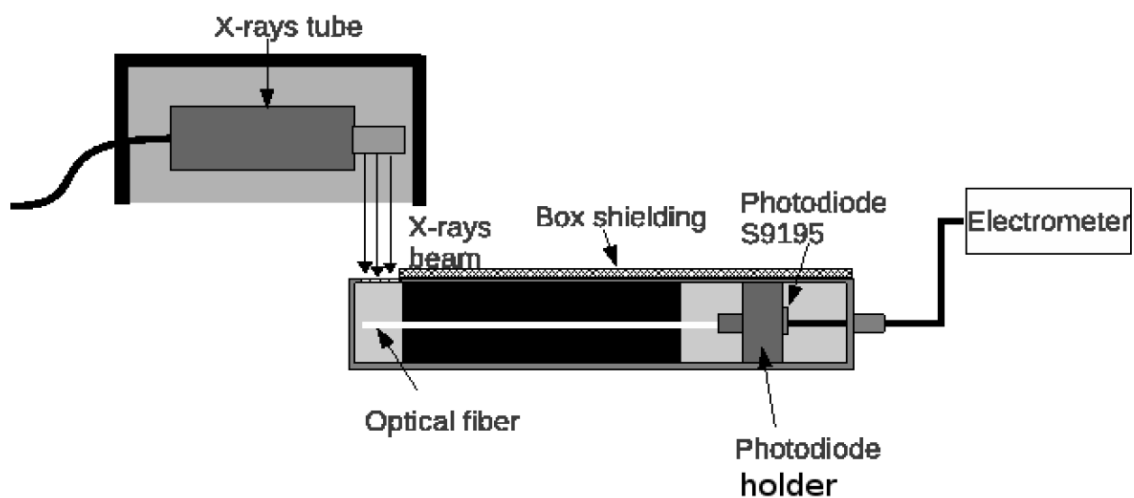

105

Figure 1: Experimental setup for testing of the photodiodes. The scintillating optical fiber is irradiated by X-rays produced by a Philips PW2184/00 tube and produces the light detected by the photodiode.

110

*II.A.2. Photodiode temperature dependence*

To use the photodiode as a photodetector in dosimetry good temperature stability is important. The relative photodiode response to visible and infrared light was measured using a special setup where a constant flux of dry air passing through the sample holder box was used to set the temperature (figure 2).

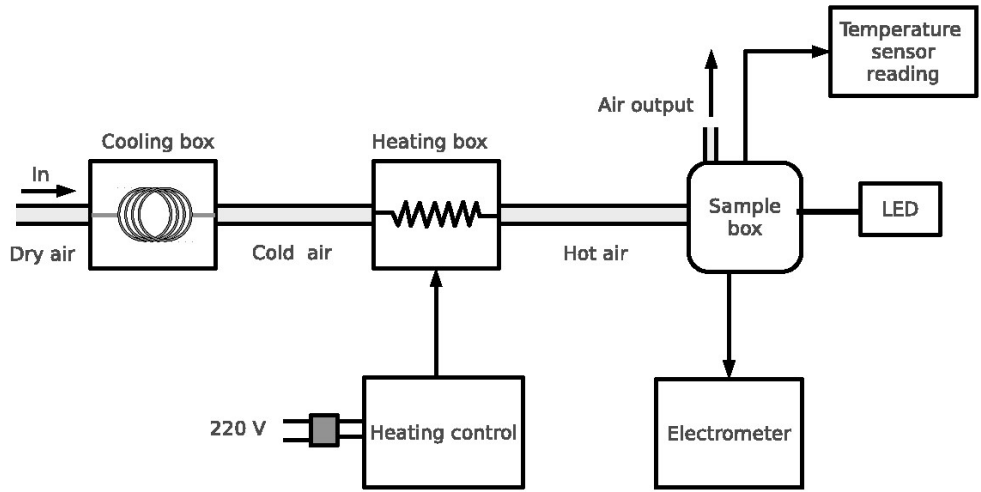

Figure 2. Experimental setup to study the temperature dependence of the photodiodes' response.

For measurement, the photodiode was placed on a plastic sample holder inside an aluminum sample-box. The sample-box wall was 10 mm thick and was wrapped in plastic insulating sheet. The box had two nozzles, one for air entry and another for air exit on the opposite wall. Inside the box the temperature was measured with a LM135[§] integrated temperature sensor with an accuracy of +-0.5 °C when calibrated at 25 °C. Dry air at room temperature was admitted into a cooling-heating circuit to get it at the desired temperature inside the sample-box. At a first stage the air was conducted through a cooper coil shape pipe placed inside ice and its temperature lowered. Then, the air passed through a heating box where an electrical resistance boosted the air temperature to higher values before entering the sample-box. A LED driven by a current source produced the light signal that was conducted by an optical fiber to the photodiode inside the sample-box. In this way the LED was always kept at constant room temperature. The signal produced in the photodiode was measured by a Standard Imaging Max4000 electrometer. Four different LED were used in the measurements and the peak wavelength of each LED was: 470 nm for blue, 565 nm for green, 660 nm for red and 930 nm for infrared.

---
[§] http://www.national.com/ds/LM/LM135.pdf

### II.A.3. Scintillating optical fiber selection

In this work, green and blue, 2 mm in diameter, scintillating optical fibers[**] from Saint-Gobain were tested. A summary of the fibers characteristics is presented in table 1.

Table 1. Characteristics of the tested optical fibers, produced by Saint-Gobain (www.detectors.saint-gobain.com/fibers.aspx).

| Type | Color | Diameter (mm) | Emission peak (nm) | Light yield (no. photons/MeV) |
|---|---|---|---|---|
| BCF-10 | Blue | 2 | 432 | 8000 |
| BCF-60 | Green | 2 | 530 | 7100 |

The same setup (figure 1) used for photodiode testing was used to measure the output signal of the fiber as function of dose. As before, 30 cm long fibers were used inside the test-box. The output signal from the photodiode was read by a Standard Imaging Max4000 electrometer. To measure the dose at the fiber position a PTW M2342-1407 ionization chamber connected to a PTW UNIDOS E electrometer was used. After removing the test-box from the X-ray beam the ionization chamber was placed exactly where the fiber end stood. In this way, for each X-ray tube accelerating potential and anode current a value of dose in air was measured at the fiber end position and a charge-dose plot was obtained for each fiber type.

---

[**] http://www.detectors.saint-gobain.com/uploadedFiles/SGdetectors/Documents/Brochures/Scintillating-Optical-Fibers-Brochure.pdf

**II.B. Fluorescent light background**

Apart from scintillation light, the beam photons are also able to produce fluorescent and Cherenkov light in the fibers. Fluorescent light production[1,2] is a process present even at low energy (few keV photons are able to produce fluorescence) while Cherenkov light production has a threshold on the secondary electron relative velocity $v/c$ given by[5]

$$\frac{v}{c} \geq \frac{1}{n} \qquad (1)$$

where $n$ is the material refractive index and c the velocity of light in vacuum. Using the relations[20] $v/c = pc/E$ where $E$ and $p$ are respectively the electron total energy and momentum, and $E = E_k + m_e c^2$ where $E_k$ is the kinetic energy and $m_e$ the electron mass, one arrives to

$$E_k \geq m_e c^2 \left( \frac{n}{\sqrt{n^2 - 1}} - 1 \right). \qquad (2)$$

For polystyrene with $n$ equal to 1.60 a threshold of 143.6 keV is obtained for the secondary electron kinetic energy and therefore the primary photon energy must be higher than this value. So, in the studied energy range, using an X-ray beam, the irreducible physical background will arise solely from fluorescent light.

170  To assess the amount of fluorescent light produced in the fibers, a clear fiber of the same material as the blue and green fibers (Saint-Gobain BCF-98) was prepared in the same way. Then, using the same setup, as explained before, one by one the signal output from the fibers inside the test-box was measured as function of the kVp. From these measurements (figure 3-left), the fluorescence contribution to the signal can be estimated by subtracting the clear fiber signal to the scintillator fiber

175  signal. The percentage of fluorescent light output ranges from 0.5 to 3% and dependents on the kVp value (figure 3- right).

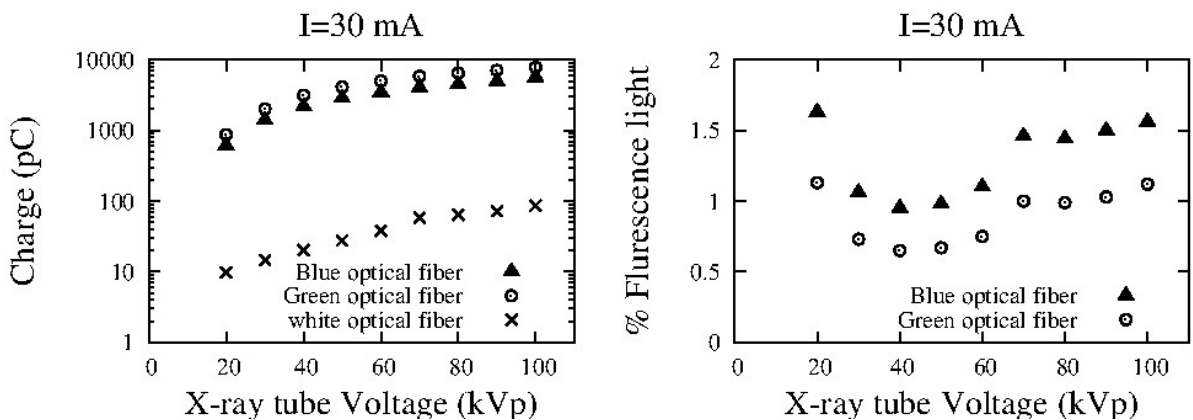

Figure 3. (left) Comparison between the signal produced by the fiber (blue, green or white/clear) as

180  function of the X-ray tube accelerating potential, (right) estimated percentage of fluorescent light.

The output measured value is affected by the collection efficiency between the fiber and the photodetector, which in turn is dependent on the fiber polish. A set of measurements with fibers cut from the same 2 m-long fiber and polish in the same way was made. An average percentage difference

185  between cuts of 10% was measured. This value contributes to an overall uncertainty up to 20 % in the relative fluorescence output.

**II.C The DosFib prototype**

190   The dosimeter prototype uses a 5 mm-long, 2 mm in diameter scintillating fiber, connected to a clear fiber (also 2 mm in diameter but with a length depending on the dosimeter application). The clear fiber is connected to a Hamamatsu S9195 photodiode. The fibers are protected from light by a thin black plastic jacket. The optical coupling between fibers and between the clear fiber and the photodiode is assured by optical grease (Saint-Gobain BC-630). The photodiode was fixed at a PVC holder which

195   ensures the clear fiber immobilization. A drawing of the dosimeter prototype is displayed in figure 4.

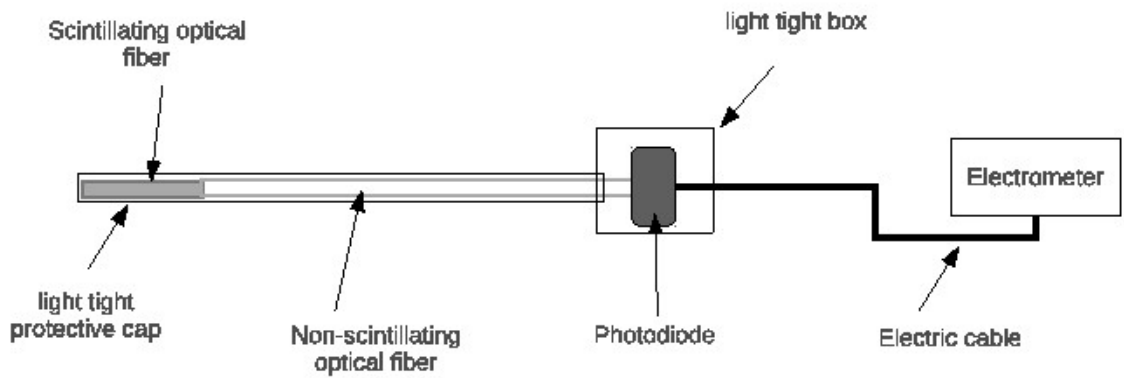

Figure 4. Drawing of the DosFib, optical fiber dosimeter prototype.

The current produced by the photodiode will depend on several factors, namely the scintillating fiber

200   light yield $y_{ph}$ (number of photons $N_{ph}$ per deposited energy $E_f$ in the fiber), the photodiode photo-sensitivity $S_d$ (current per incident power in A/W), acceptance efficiency $\varepsilon_a$ and coupling efficiencies $\varepsilon_1$, $\varepsilon_2$ between fibers and clear fiber to photodiode. A simple model can give the order of magnitude of the device sensitivity as follows.

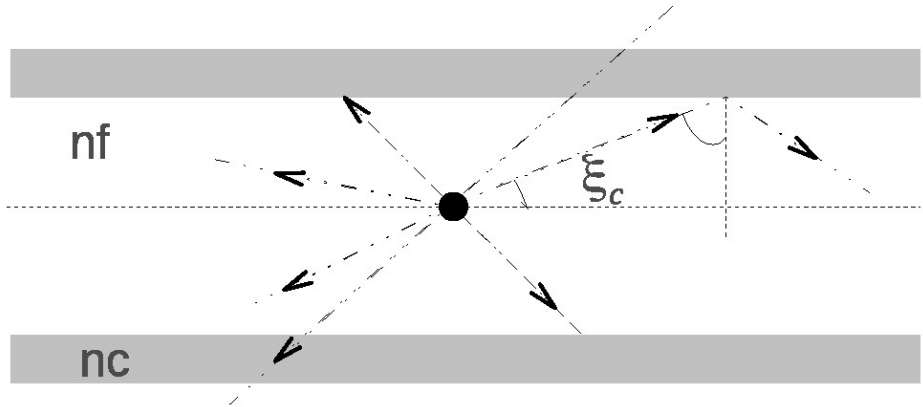

Figure 5. Light paths inside the optical fiber.

For each deposited MeV in the fiber an average number $N_{ph}$ of scintillation photons is produced, but not all of them will exit from the fiber end. Neglecting light attenuation in the scintillator (a reasonable assumption for a small scintillator, but not true for the clear fiber) only photon conducted by total reflection inside the fibers' core will have a chance to exit (figure 5). If $n_f$ and $n_c$ are respectively the fibers' core and cladding refractive index from Snell's law[21] one gets for the critical angle $\theta_c$ as

$$n_f \sin(\theta_c) = n_c. \tag{3}$$

From figure 5 the critical emission angle $\xi_c$ for light conduction by total reflection is $\xi_c = \dfrac{\pi}{2} - \theta_c$ and we thus have

$$\cos(\xi_c) = \sin(\theta_c) = \frac{n_c}{n_f}. \tag{4}$$

The fraction of photons exiting the fiber by one of its extremities will be[22,23]

$$\varepsilon_a = \frac{1}{4\pi} \int_{\cos\xi_c}^{1} d\Omega \qquad (5)$$

or integrating

$$\varepsilon_a = \frac{1}{2}\left(1 - \frac{n_c}{n_f}\right). \qquad (6)$$

For polystyrene fibers with acrylic cladding we have $n_f=1.60$ and $n_c=1.49$ and one gets $\varepsilon_a = 0.034$. The coupling efficiencies depend on the quality of the fiber top polish and are difficult to evaluate. Thus we will assume a conservative value of 0.5 as suggested by Beddar et al (Ref. 22) for both, scintillator-clear fiber $\varepsilon_1$, and clear fiber-photodiode $\varepsilon_2$ efficiencies. Assuming an exponential attenuation in the clear fiber with attenuation length $L=2.2$ m the fraction of light surviving $x$ meters of fiber will be $f \approx e^{-x/L}$ and for a 1m-long clear fiber we get $f$ equal to 0.635. The total light collection efficiency will then be $\varepsilon = \varepsilon_a \varepsilon_1 \varepsilon_2 f$.

We take the fiber dosimeter sensitivity as the produced charge $\Delta q$ in the photodiode per deposited dose $\Delta D$ in the scintillator $S = \frac{\Delta q}{\Delta D}$. The produced charge $\Delta q$ in the photodiode is given by $S \times E_{ph}$ where the energy from visible photons $E_{ph}$ impinging on the photodiode is given by

$$E_{ph} = N_{ph} \frac{hc}{\lambda},  \quad (7)$$

being $hc/\lambda$ the energy of each photon, $\lambda$ its wavelength, $h$ the Planck's constant and $c$ the light velocity. As defined above the number of photons produced in the fiber is given by $N_{ph} = y_{ph} E_f$, while the deposited energy in the fiber can be related to the dose D through $E_f = D\rho V$ where $\rho$ is the fiber density and $V$ its volume. So, for the dosimeter sensitivity one finally arrives to

$$S = S_d y_{ph} \rho V \varepsilon \frac{hc}{\lambda}. \quad (8)$$

According to the Saint-Gobain datasheet the blue BCF-10 fibers have an emission peak of 432 nm and a photon yield of $8000/1.6 \times 10^{-13}$ photons/J and a density of 1050 kg/m³. A dosimeter with a 5 mm-long and 2 mm in diameter blue scintillating fiber coupled to a 1 m-long BCF-98 clear fiber read by an S9195 photodiode with photo-sensitivity of 0.28 A/W will have a sensitivity of about 0.6 nC/Gy. This value is in close agreement with the obtained experimental result (see figure 10), confirming the reasonableness of the assumptions made.

## III. RESULTS

### III.A. Dosimeter component selection

#### *III.A.1. Photodetector selection*

270  For each photodiode, measurements were made for two different tube potentials (20, and 100 kVp) and for three anode currents: 10, 20 and 30 mA. Figure 6 presents the obtained results for the photodiode output charge (collected in 30 s) as a function of the tube current for three tube accelerating potentials, are presented. These results show the S9195 better performance on what concerns the signal output for a majority of cases and accordingly that photodiode was our choice.

275

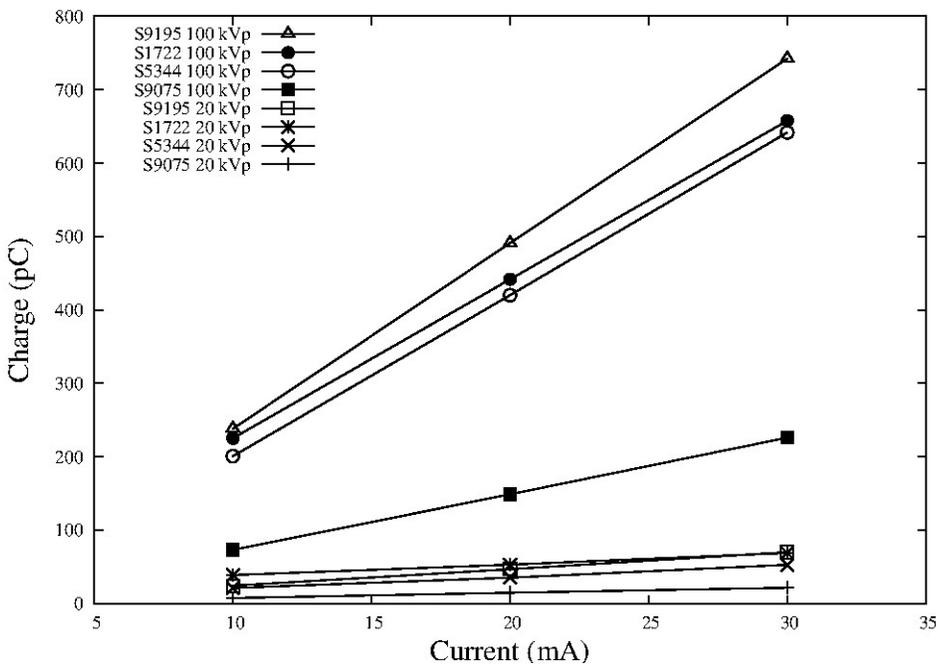

Figure 6. Photodiode charge response as function of anode current in the X-ray tube for two different accelerating potentials.

280

### III.A.2. Photodiode temperature dependence

The photodiodes output signal was measured for temperatures between 0 and 50 °C. Output values
285 were normalized to the room temperature (20° C) value for each exciting LED. Measurements were made for the full set of photodiodes, but since results are similar only the values obtained for the S9195 photodiode are presented in figure 7. For this photodiode a variation no greater than 2% in the output signal is observed in the temperature range between 0 and 50 °C.

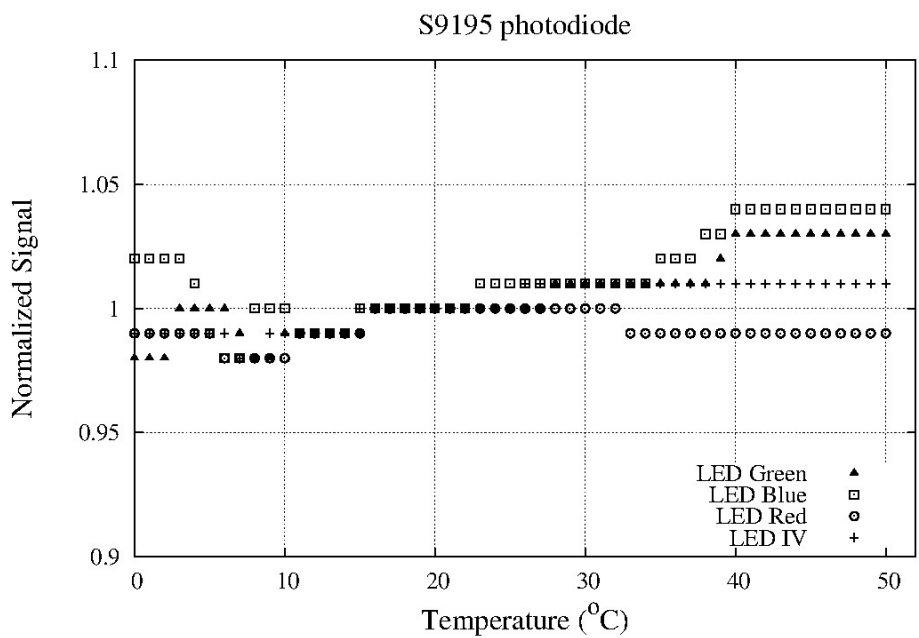

290

Figure 7. Normalized photodiode response as function of temperature when irradiated with light from different LEDs.

The rather good temperature independence and stability is a plus relative to the use of a PMT as
295 photodetector. For instance, the typical PMT used for dosimetry applications,[2,14] (R647 from

Hamamatsu[††]) has a decrease in sensitivity of about 20% in the 0 to 40 °C range. The PMT temperature dependence introduces the need of larger calibration factor corrections relative to the ones needed by photodiodes.

### III.A.3. Scintillating optical fiber selection

In figures 8-left and 8-right the obtained fiber response (in pC) as a function of the dose (in Gy) is displayed for several accelerating tube potential (kVp). No extra filtration of the X-ray beam was done for these tests.

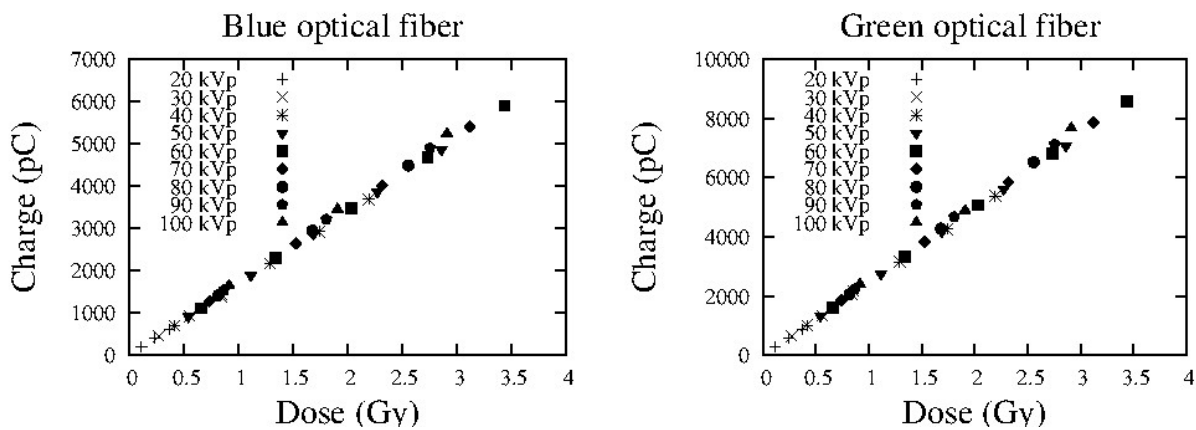

Figure 8. Photodiode signal, when coupled to a blue (left) or green (right) optical fiber as a function of dose, measured with an ionization chamber for different X-ray tube accelerating potentials.

To assess the data linearity, linear fits were performed using the data for each kVp and the difference between the measured and fitted value. For a majority of values a difference less then 0.3% was observed.

---

[††] http://www-d0.fnal.gov/icd/docs/r647_pmt_spec.pdf

The sensitivity $S = \frac{\Delta q}{\Delta D}$ as function of the tube kVp is presented in figure 9. For the sake of comparison, sensitivity values have been normalized to the 100 kVp sensitivity value. The behavior of both green and blue fiber sensitivity curves is almost identical, with a steady increase of about 12% from 30 to 100 kVp.

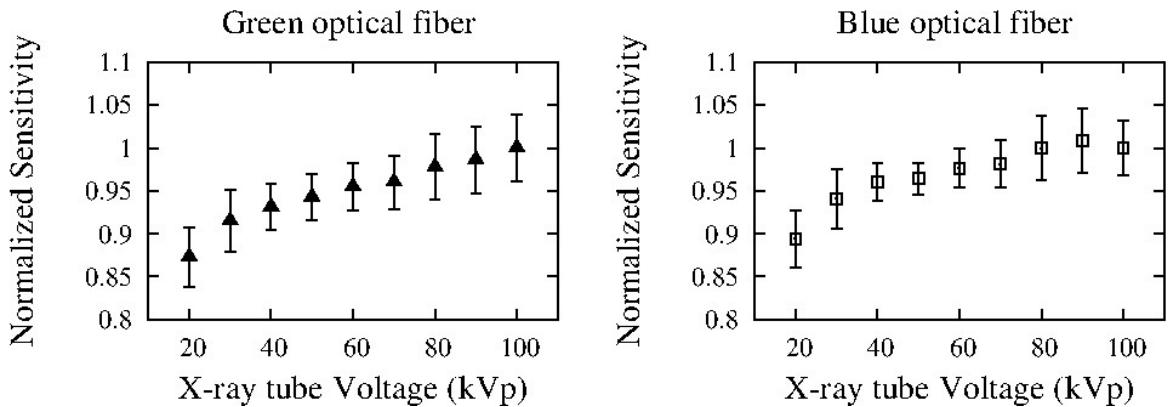

Figure 9. Normalized sensitivity versus kVp for BCF-60 (green) and BCF-10 (blue) fibers.

**III.B. DosFib response to X-rays**

The DosFib prototype was tested under an X-ray beam produced by the same X-ray tube as before. To perform the study under more realistic conditions the dosimeter was placed inside an acrylic phantom at a depth of 3 mm. For these tests a lead collimator with a 2.3 cm aperture was used at the X-ray tube exit window, but no further filtration was considered. As before, the dose at the scintillator position was acquired with a PTW M2342-1407 ionization chamber, now also placed inside an acrylic phantom at the same depth as the fiber. The dosimeter response (in pC) as function of dose for several X-ray tube kVp is presented in figures 10-left (blue fibers) and 10-right (green fibers).

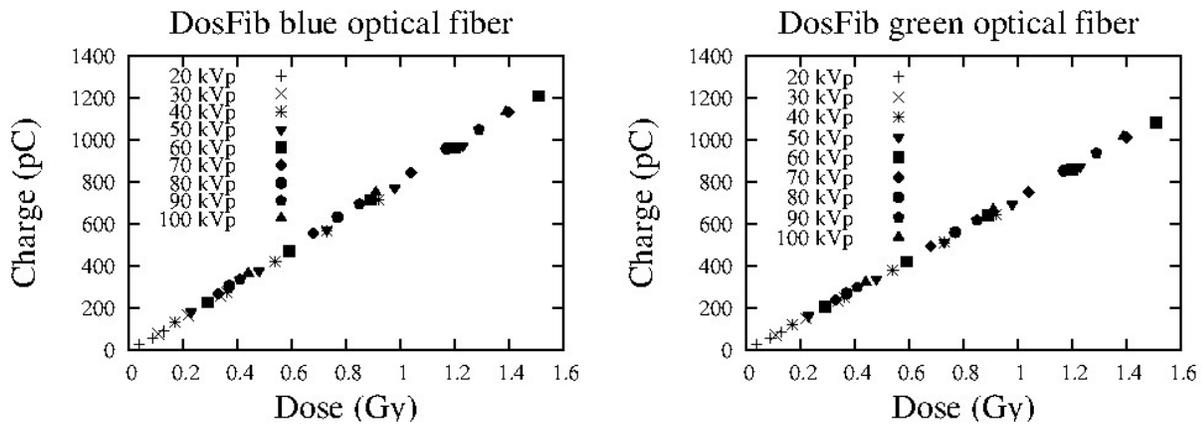

330     Figure 10. The DosFib response as function of dose, using a blue (left) or green (right) scintillating optical fiber.

From these curves, sensitivities can be derived as function of the X-ray tube kVp, for both blue and green fibers (figure 11).

335

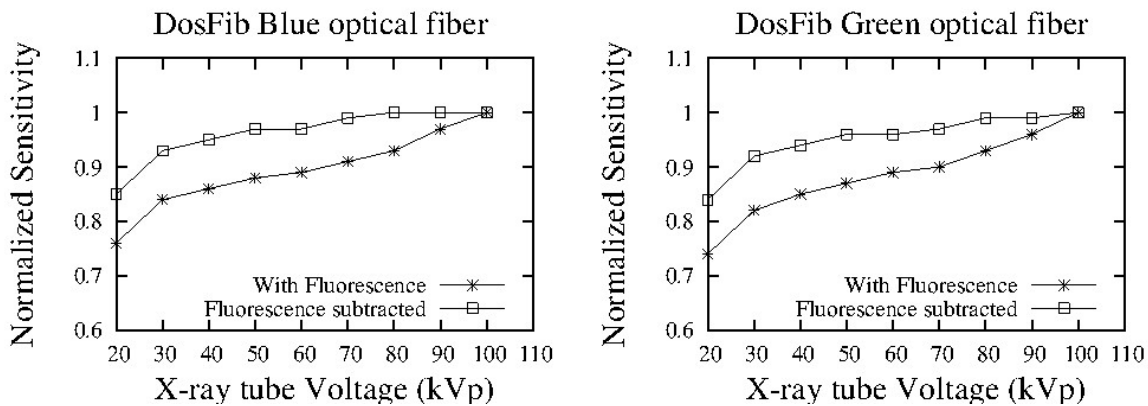

Figure 11. DosFib normalized sensitivity as function of X-ray tube accelerating potential using a blue (left) or green (right) scintillating optical fiber. Sensitivity curves after subtraction of fluorescent light are also shown (see text for details).

340

Again, these sensitivity values are affected by the production of fluorescent light in the scintillator and in the clear fiber light-guide. Fluorescent light production can be measured if the scintillator is removed from the dosimeter extremity and solely the clear fiber response is acquired. The fluorescent light contribution can then be subtracted from the overall dosimeter response. Sensitivity curves were then obtained using these fluorescence-free values, as also seen in figure 11. The fluorescent light removal improves the sensitivity curve in the way to make it more energy independent. This result confirms fluorescent light as a major cause for the energy dependence of this type of dosimeter at low energies.

**III.C. DosFib Clinical tests**

High Dose Rate (HDR) $^{192}$Ir sources are widely used nowadays for brachytherapy treatments. $^{192}$Ir is a beta emitter with 74 day half-life and complex gamma and X-ray photon spectrum[24] with mean energy of 370 keV. Tests of the DosFib were conducted with a clinical $^{192}$Ir source at the Hospital de Santa Maria in Lisbon. The radioactive source was housed in a Nucletron microSelectron afterloading system. A special water/acrylic phantom was developed for these measurements (figure 12). The phantom consisted of an acrylic bench placed inside an acrylic box with 30 cm on each side to be filled with water.

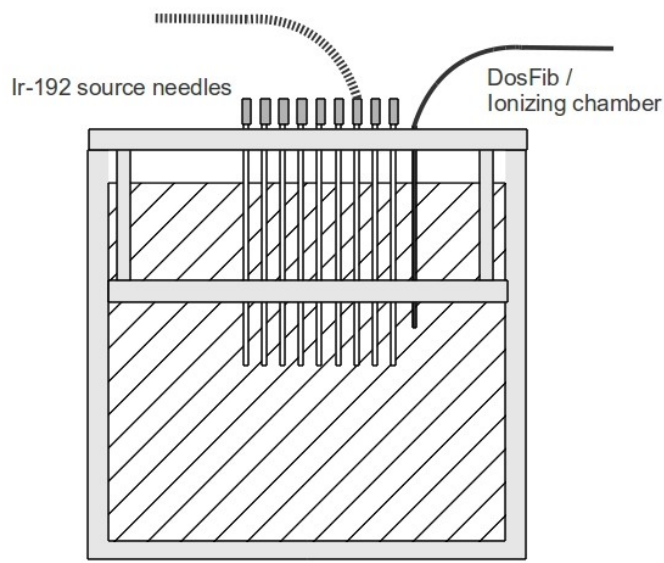

Figure 12. Acrylic and water phantom for DosFib tests at the hospital. The $^{192}$Ir source leaves the afterloader running inside a flexible cable into one of the plastic needles.

360

Two rows of holes were drilled on the acrylic bench. The first position of the first row was reserved for the catheter bearing the DosFib, while the first position of the second row was reserved for placing a 0.3 cm3 PTW 31003 ionization chamber. At the other holes plastic catheters were inserted and the $^{192}$Ir source could move freely within each one of them. In this way a set of measurements at fixed

365 dosimeter-source distance could be made in a reproducible way. To find the vertical source measurement position, a set of measurements were performed at different source positions moving the source inside one of the catheters. The optimal vertical measurement position was then chosen at the maximum obtained signal position (figure 13).

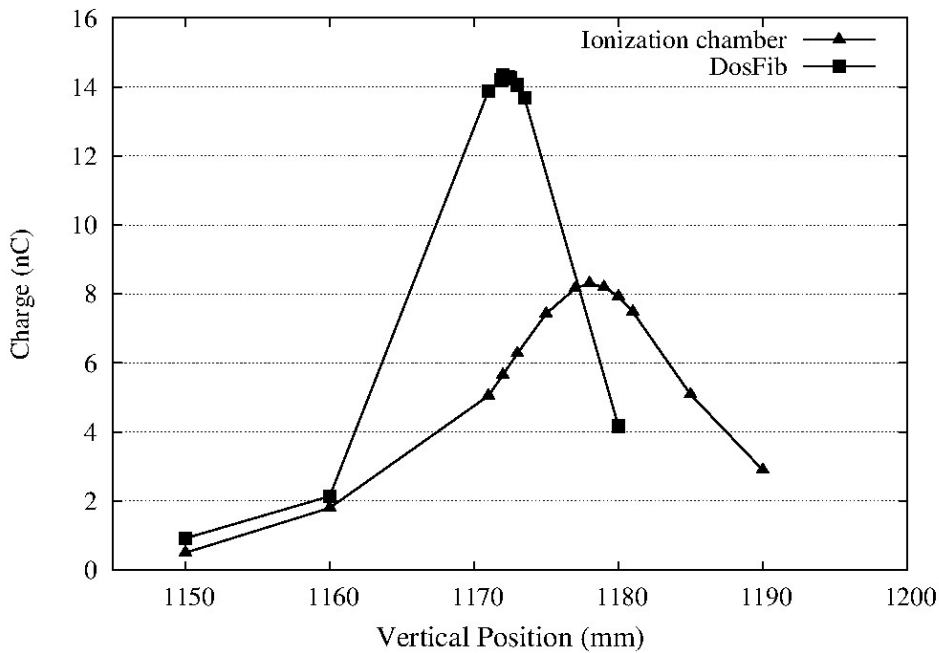

370

Figure 13. Determination of the measuring point. The vertical position of the $^{192}$Ir source was adjusted to obtain the maximum signal.

Since a step-by-step motor drove the source, it could be, in a reproducible way, re-positioned at this

375 vertical position when placed inside any other catheter. As can be seen from figure 13, the DosFib and ionization chamber maximum signal position values do not match, because there was a poor vertical accuracy in the detectors placement. Thus, this procedure ensured the correct source-detector alignment.

380  Several measurements were made with the DosFib and the PTW ionization chamber for distances $d$ source-dosimeter of clinical relevance (figure 14). For comparison purposes the measurements in figure 14 were further normalized to unit at the source-dosimeter distance of 1 cm.

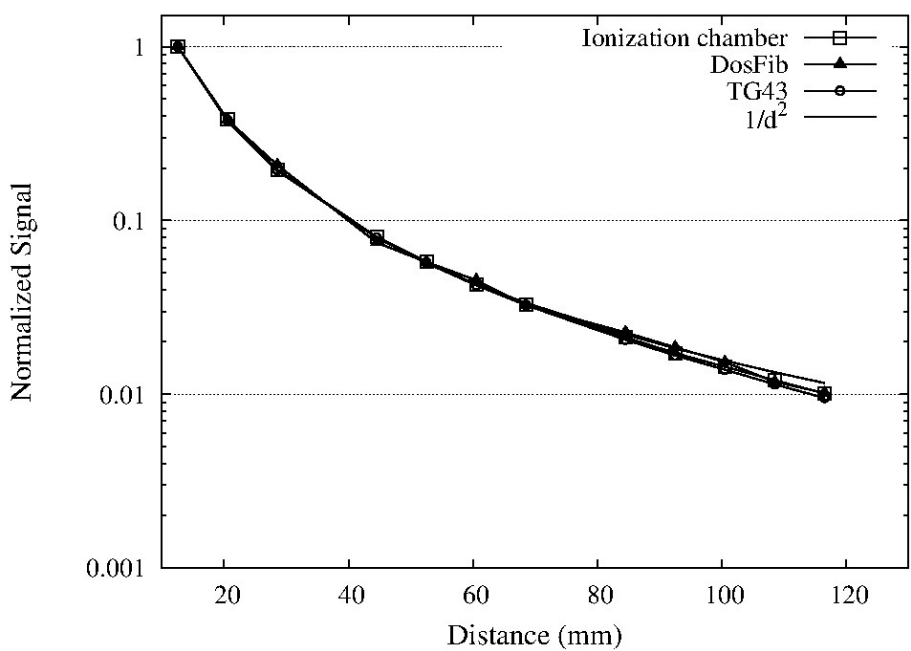

385  Figure 14. DosFib and ionization chamber normalized response as functions of source-dosimeter distance for a $^{192}$Ir source. The comparison with the TG43 result prevision[25] for the dose dependence with distance is made as well as the comparison with the attenuation curve due to the squared inverse distance law.

390  A $1/d^2$ tendency curve was been added to the plot, as well as the TG43 prevision[25] for the dose dependence on the distance. As can be seen from the figure an almost match is obtained between the

DosFib and ionization chamber response, closely following the TG43 curve. The difference between the DosFib and ionization chamber normalized signals is less than 5% for distances up to 10 cm. Since the fiber tends to bend in the end we estimate that some of the difference between the ionization

395 chamber and the DosFib is due to placement mismatch. To overcome this problem the construction of a full acrylic phantom is foreseen. The agreement between the DosFib and ionization chamber curves show that, despite the fluorescent light contribution (and for a $^{192}$Ir source, also Cherenkov light), a calibration of the DosFib and accurate dose measurements are possible.

400 **IV. CONCLUSIONS**

The DosFib prototype has demonstrated the suitability of photodiodes as photodetectors for optical fiber dosimetry. They have a number of advantages over traditional PMTs, starting with the ability to work with no bias. The chosen photodiode (Hamamatsu S9195) for this work has almost flat temperature dependence over the practical working range (15 to 35 ºC). The DosFib uses the BCF-10

405 blue fiber from Saint-Gobain as the scintillating element, coupled to a plastic clear fiber. The total amount of fluorescent light, mainly produced in the clear fiber, amounts up to 5% in practical measurement situations, but is not enough to compromise the prototype as a dosimeter, if a proper calibration is made. Concerning the dosimeter energy dependence, a 12% sensitivity variation was measured in the range 20 to 100 kVp, but a variation of only 6% between 50 and 100 kVp. The

410 obtained sensitivities, in the range of 0.6 to 0.7 nC/Gy, allow the direct use of an electrometer as the measuring device, greatly simplifying the measuring circuit. Finally, the DosFib prototype has been tested in a water phantom and the depth dose curve obtained with a $^{192}$Ir clinical brachytherapy source is identical to the one obtained with an ionizing chamber in the same conditions.

415

**Acknowledgments**

We want to thank to Laboratório de Instrumentação e Física Experimental de Partículas (LIP) for the financial support of this work and in particularly the LIP mechanical workshops for the construction of the water phantom container. We thank Prof. Maria Luisa Carvalho for the use of the X-ray facility at the Lisbon University. Finally we acknowledge the Hospital de Santa Maria in Lisbon and Dr. Carlos Jesus, for the opportunity to perform in-vitro tests in the hospital radiotherapy service.

500